# Optimal Column-Based Low-Rank Matrix Reconstruction [*]

Venkatesan Guruswami[†]    Ali Kemal Sinop[‡]

Computer Science Department
Carnegie Mellon University
Pittsburgh, PA 15213.

**Abstract**

We prove that for any real-valued matrix $X \in \mathbb{R}^{m \times n}$, and positive integers $r \geqslant k$, there is a subset of $r$ columns of $X$ such that projecting $X$ onto their span gives a $\sqrt{\frac{r+1}{r-k+1}}$-approximation to best rank-$k$ approximation of $X$ in Frobenius norm. We show that the trade-off we achieve between the number of columns and the approximation ratio is optimal up to lower order terms. Furthermore, there is a deterministic algorithm to find such a subset of columns that runs in $O(rnm^\omega \log m)$ arithmetic operations where $\omega$ is the exponent of matrix multiplication. We also give a faster randomized algorithm that runs in $O(rnm^2)$ arithmetic operations.

## 1  Introduction

Given a matrix $X \in \mathbb{R}^{m \times n}$ and a positive integer $k < n$, the best rank-$k$ approximation to $X$ is given by top $k$ singular vectors of $X$:

$$X_{(k)} = \sum_{i=1}^{k} \sqrt{\sigma_i} u_i v_i^T$$

where $\sigma_1 \geqslant \sigma_2 \geqslant \ldots \geqslant \sigma_n \geqslant 0$ are the eigenvalues of $X^T X$, and $u_i$ (resp. $v_i$) are the associated left (resp. right) singular vectors for each singular value $\sqrt{\sigma_i}$. Furthermore $X_{(k)}$ can be computed in time $O(\min(n,m)mn)$-time using Singular Value Decomposition (SVD).

One related question that has received considerable attention in recent years is choosing $r$ *columns* of $X$, for some input parameter $r \geqslant k$, whose span approximates $X$ as nearly as well as $X_{(k)}$. In other words, we would like to relate

$$\min_{C \in \binom{[n]}{r}} \|X - X_C^\Pi X\|_\xi = \min_{C \in \binom{[n]}{r}} \|X_C^\perp X\|_\xi$$

to $\|X - X_{(k)}\|_\xi$ for some norm $\xi$, and efficiently find a subset $C$ of $r$ columns coming close to this bound. Here $X_C$ denotes matrix formed by columns of $X$ corresponding to $C$ and $X_C^\Pi$ (resp. $X_C^\perp$) is the projection matrix onto $X_C$ (resp. onto null space of $X_C$).

This basic problem seems well-motivated in various application settings. For example, this problem has applications in data sets arising from document classification problems, face recognition tasks, and so on, where it is important to pick a subset of features that are dominant (and it is not appropriate to work with linear combinations of features output by usual dimension reduction techniques like random projection or singular value decomposition). We refer the reader to Mahoney and Drineas [10] for comparisons of SVD and column selection on experimental data.

Our interest in this problem stemmed from our own work on improved approximation algorithms using certain Semidefinite Programming relaxations from the so-called Lasserre Hierarchy [8]. The analysis of our algorithms relied on bounded quantities such as $\min_{C \in \binom{[n]}{r}} \|X_C^\perp X\|_F$ for the Frobenius norm. In this application, the running time is exponential in $r$, where $r$ is the number of columns one has to choose from $X$ to approximate $X$ in Frobenius norm as close to $X_{(k)}$ as possible. Thus finding the optimal dependence between $r$ and $k$ was a question of natural significance.

Our main results in this paper are the following two theorems. We are able to get the best known dependence between $r$ and $k$, show its optimality up to lower order terms, and achieve this with an efficient deterministic algorithm (Theorem 1.1). This answers one of the open questions mentioned in [1]. We are also able to give a more efficient randomized algorithm, via a faster implementation of exact volume sampling (Theorem 1.2). The deterministic algorithm of Theorem 1.1 is a derandomization of the volume sampling algorithm

[*]To appear in Proceedings of the $23^{rd}$ Annual ACM-SIAM Symposium on Discrete Algorithms, 2012.

[†]Supported in part by a Packard Fellowship and NSF CCF-0963975. Email: `guruswami@cmu.edu`

[‡]Supported in part by NSF grants CCF-0953155, CCF-1115525, and MSR-CMU Center for Computational Thinking. Email: `asinop@cs.cmu.edu`

via conditional expectations [4].

THEOREM 1.1. *Given $X \in \mathbb{R}^{m \times n}$, and positive integers $k \leqslant r$, one can find a set $C$ of $r$ columns, deterministically using at most $O(rnm^\omega \log m)$ many arithmetic operations (where $\omega$ is the exponent of matrix multiplication), such that*

$$(1.1) \qquad \|X - X_C^\Pi X\|_F^2 \leqslant \frac{r+1}{r+1-k}\|X - X_{(k)}\|_F^2 \; .$$

*Furthermore, for any $r = o(n)$, this bound is tight up to lower order terms.*

THEOREM 1.2. *Given a matrix $X \in \mathbb{R}^{m \times n}$, $m \leqslant n$, and $r \geqslant 1$, there is an algorithm* `Volume-Sample` *that samples a subset of $r$ columns of $X$, $C \in \binom{[n]}{r}$, with probability $\frac{|X_C^T X_C|}{\sum_{T \in \binom{[n]}{r}} |X_T^T X_T|}$ using at most $O\left(rnm^2\right)$ arithmetic operations. For every $k \leqslant r$, the subset $C$ returned by* `Volume-Sample` *satisfies*

$$\mathbb{E}_C\left[\|X - X_C^\Pi X\|_F^2\right] \leqslant \frac{r+1}{r+1-k}\|X - X_{(k)}\|_F^2 \; .$$

Note that $\|X - X_C^\Pi X\|_F^2 = \|X_C^\perp X\|_F^2 = \mathsf{Tr}(X^T X_C^\perp X)$. Henceforth in this paper, we will use the Trace notation.

**1.1 Relation to previous work** The first algorithm for $k$-column matrix reconstruction was given in a seminal paper by Frieze, Kannan and Vempala [7], where they presented a randomized algorithm to find $\text{poly}(k, 1/\varepsilon, 1/\delta)$ columns that achieve an *additive* error of $\varepsilon\|X\|_F$.

Subsequent works concentrated on removing the additive factor and getting multiplicative (or relative error) guarantees, and improving the dependence between $r$ and $k$ to get a desired relative error. Some of these works are mentioned in Figure 1. In the table, $r$ is the number of columns needed so as to obtain the given approximation ratio, defined as $\mathsf{Tr}(X^T X_C^\perp X)/\|X - X_{(k)}\|_F^2$.

To briefly place our result in context, let us mention the known existential bounds on the relation between $r$, $k$, and the ratio achieved. Deshpande *et al* [5] prove the existence of $k$ columns achieving a ratio $k+1$, and also show that this is best possible up to lower order terms. Deshpande and Vempala [6] prove that for small $\varepsilon > 0$, there exists a matrix $M$ for which the best error achieved by a rank-$k$ matrix, whose columns are restricted to belong to the span of $r \geqslant k/\varepsilon$ columns of $M$, is at least $1 + \varepsilon - o(1)$ times the best rank-$k$ approximation.[1]

Until recently, even the best existential bound to achieve $(1 + \varepsilon)$ approximation was super-linear in $k$. In an independent and concurrent work, Boutsidis, Drineas, and Magdon-Ismail [1] showed a bound of $r \approx k + \frac{2k}{\varepsilon}$ along with a randomized algorithm to find such a subset of columns.[2] Our main result proves that $k/\varepsilon + k - 1$ columns are sufficient, and further those columns can be found in *deterministic* polynomial time.

The $(1 + \varepsilon)$ approximation achieved in [1] holds in the restricted model (in which the above-mentioned $k/\varepsilon$ lower bound of [6] applies) where one must find a *rank-$k$ approximation matrix* contained in the span of the chosen $r$ columns, whereas our approximating matrix uses the full span of the chosen columns. So our results and [1] are incomparable in this respect. We stress though that even allowing for full column span, no bounds on $r$ which were linear in $k$ were known till recently, for achieving say a factor 2 approximation. Further, we extend the lower bound in [6] to show that even allowing for full column span, $r = k/\varepsilon$ columns are needed for a factor $(1 + \varepsilon - o(1))$ approximation.

Note that our result gives the optimal $(k+1)$ factor approximation (taking $\varepsilon = k$) for $r = k$, and for $\varepsilon \to 0$, the near-optimal $(1+\varepsilon)$ factor for $r \approx k/\varepsilon$, in a uniform way. As for the algorithmic claim, recently Deshpande and Rademacher [4] gave an efficient implementation of volume sampling and a deterministic algorithm to find a set $k$ columns with approximation ratio $k + 1$, thus matching the bound of [5] algorithmically. We simply bound the ratio achieved by this algorithm when it is allowed to pick $r > k$ columns. In other words, the algorithmic part of Theorem 1.1 follows from [4], given our combinatorial bound.

Prior to our work, the fastest algorithm known for *exact* volume sampling was given in [4] using $O(rnm^\omega \log m)$ arithmetic operations. We give an asymptotically faster sampling algorithm, by using binary search to pick the lowest index column in the sampled set with the correct marginal probability, and then recursing to sample the remaining $r - 1$ columns.

**1.2 Our Techniques** Our proof is based on the following bound:

$$(1.2) \qquad \min_{C \in \binom{[n]}{r}} \mathsf{Tr}(X^T X_C^\perp X) \leqslant (r+1)\frac{\mathbb{S}_{r+1}(\sigma)}{\mathbb{S}_r(\sigma)}$$

$$= \mathbb{E}_{\mathbf{C} \sim \mathcal{C}_r(X)}\left[\mathsf{Tr}(X^T X_{\mathbf{C}}^\perp X)\right]$$

---

[1] Although in [6] the lower bound is stated as $1 + \frac{\varepsilon}{2} - o(1)$, the actual lower bound they prove is stronger and equals $1 + \varepsilon - o(1)$.

[2] The theorem statement in [1] mentions the weaker bound $r \leqslant 10k/\varepsilon$, but the sharper bound is given at the end of Section 4 of the paper.

| Paper | $r$ | Ratio | Running Time | Deterministic |
|---|---|---|---|---|
| This work | $k + \frac{k}{\varepsilon} - 1$ | $1+\varepsilon$ | $O(rnm^\omega \log m)$ | Yes |
| This work | $k + \frac{k}{\varepsilon} - 1$ | $1+\varepsilon$ | $O(rnm^2)$ | No |
| [1] | $\frac{2k}{\varepsilon}$ | $1+\varepsilon$ | $O(k\varepsilon^{-1}nm + k^3\varepsilon^{-2/3}n)$ | No |
| [4] | $k$ | $k+1$ | $O(knm^\omega \log m)$ | Yes |
| [3] | $O(k^2 \log k + k\varepsilon^{-1})$ | $1+\varepsilon$ | $O(k^2 mn \log k)$ | Yes |
| [12] | $O(k \log k + k\varepsilon^{-1})$ | $1+\varepsilon$ | $O((k \log k + k\varepsilon^{-1})mn + (k \log k + k\varepsilon^{-1})^2)(m+n)$ | No |
| [6] | $O(k^2 \log k + k\varepsilon^{-1})$ | $1+\varepsilon$ | $O(k^2 mn \log k)$ | No |

Figure 1: Performance and running time of various column selection algorithms

where $\mathbf{C} \sim \mathcal{C}_r(X)$ denotes sampling $C$ with probability proportional to determinant of $X_C^T X_C$, $|X_C^T X_C|$, and $\mathbb{S}_r(\sigma)$ is the $r$'th symmetric function of $\sigma_1, \sigma_2, \ldots, \sigma_n$. The bound (1.2) already appears in the work of Deshpande *et al.* [5] where sampling from $\mathcal{C}_r(X)$ is called "volume sampling."

Our main technical contribution is to use the *Schur-concavity* of $\frac{\mathbb{S}_{r+1}(\sigma)}{\mathbb{S}_r(\sigma)}$ and theory of majorization [11] to bound $\frac{\mathbb{S}_{r+1}(\sigma)}{\mathbb{S}_r(\sigma)}$ in terms of $\sum_{i \geq k+1} \sigma_i$. At an intuitive level, the ratio $\frac{\mathbb{S}_{r+1}(\sigma)}{\mathbb{S}_r(\sigma)}$ should be larger when $\{\sigma_i\}_{i=1}^n$ is more "uniform." Majorization and Schur-concavity allow us to turn this intuition into a precise and formal statement. This leads us to the inequality

$$(1.3) \qquad (r+1)\frac{\mathbb{S}_{r+1}(\sigma)}{\mathbb{S}_r(\sigma)} \leq \frac{r+1}{r+1-k} \sum_{i>k} \sigma_i ,$$

which together with $\|X - X_{(k)}\|_F^2 = \sum_{i>k} \sigma_i$ and (1.2) yields the claimed bound (1.1). For the nearly matching lower bound, we prove that for the construction given in [6], the lower bound on approximation ratio holds even in the unrestricted model where the full column span of the $r$ columns is allowed; this analysis appears in Section 6.

As for the algorithm, Deshpande and Rademacher [4] used the method of conditional expectations to find $C \in \binom{[n]}{r}$ satisfying $\mathsf{Tr}(X^T X_C^\perp X) \leq \mathbb{E}_{\mathbf{C} \sim \mathcal{C}_r(X)}\left[\mathsf{Tr}(X^T X_\mathbf{C}^\perp X)\right]$ deterministically using $O(rnm^\omega \log m)$ operations. Together with our bound (1.3), this implies a deterministic algorithm achieving a $\frac{r+1}{r+1-k}$ ratio. In light of this, we do not discuss the deterministic part any further in this paper, and focus on proving and (1.3) and (1.2), which we do in Sections 3 and 4 respectively. Our more efficient volume sampling algorithm is described in Section 5. The proof of our lower bound is presented in Section 6.

## 2 Preliminaries and Notation

For any positive integer $n$, we use $[n] \triangleq \{i \in \mathbb{N} : i \leq n\}$ to denote the set of positive integers smaller than or equal to $n$. We will use $\binom{A}{k}$ to denote the $k$-subsets of $A$.

Given real vector $a = (a_i)_{i=1}^n \in \mathbb{R}^n$, we will use $a\uparrow_i$ (resp. $a\downarrow_i$) to denote the $i^{th}$ smallest (resp. largest) element of $\{a_i\}_i$.

We say $a = (a_i)_{i=1}^n \in \mathbb{R}^n$ *majorizes* $b = (b_i)_{i=1}^n \in \mathbb{R}^n$ if for all $j \in [n]$, $\sum_{j' \leq j} a\downarrow_{j'} \geq \sum_{j' \leq j} b\downarrow_{j'}$ and $\sum_j a_j = \sum_j b_j$. We denote this relation by $a \succ b$.

OBSERVATION 2.1. *For any non-negative vector $a \in \mathbb{R}^n \geq 0$, the following holds:*

$$(1, 0, \ldots, 0) \succ \frac{1}{\sum_i a_i} a \succ \left(\frac{1}{n}, \frac{1}{n}, \ldots, \frac{1}{n}\right)$$

DEFINITION 2.1. *A function $F : \mathbb{R}^n \to \mathbb{R}$ is called Schur-concave if whenever $a \in \mathbb{R}^n$ majorizes $b \in \mathbb{R}^n$, $a \succ b$, $F(a) \leq F(b)$.*

DEFINITION 2.2. (SYMMETRIC POLYNOMIALS) *For a given $\sigma = (\sigma_1, \ldots, \sigma_n) \in \mathbb{R}^n$, let $\mathbb{S}_r(\sigma)$ denote the $r^{th}$ symmetric polynomial:*

$$\mathbb{S}_r(\sigma) \triangleq \sum_{S \in \binom{[n]}{r}} \prod_{i \in S} \sigma_i.$$

*Likewise, for a given square matrix $A \in \mathbb{R}^{m \times m}$, $\mathbb{S}_r(A)$ is defined as*

$$\mathbb{S}_r(A) = \sum_{U \in \binom{[m]}{r}} |A_{U|U}| ,$$

*where $A_{U|U}$ is the minor of $A$ corresponding to columns and rows in $U$.*

LEMMA 2.1. *If $A \in \mathbb{R}^{m \times m}$ has eigenvalues $\{\sigma_i\}$, then $\mathbb{S}_r(A) = \mathbb{S}_r(\sigma)$.*

*Proof.* The coefficient of $x^{m-r}$ in $\prod_i(\sigma_i - x)$ equals $(-1)^{m-r}\mathbb{S}_r(\sigma)$. Similarly $(-1)^{m-r}\mathbb{S}_r(A)$ is the coefficient of $x^{m-r}$ in $|-xI + A|$. Now, note that $|-xI + A| = \prod_i(\sigma_i - x)$.

Given a matrix $X \in \mathbb{R}^{m \times n}$ and $i \in [n]$, we use $X_i$ to denote $i^{th}$ column of $X$. Similarly given a subset of columns, $C \subseteq [n]$, we use $X_C$ to denote the matrix formed by columns from $C$, $X_C = (X_i)_{i \in C}$. Also we will let $X^\Pi$ and $X^\perp$ be the projection matrix onto range and null space of $X$ respectively.

For any square matrix $A \in \mathbb{R}^{m \times m}$, we will use $|A|$ to denote the determinant of $A$, $\mathsf{Tr}(A)$ to denote trace of $A$ and $\sigma_i(A)$ to denote the $i^{th}$ largest eigenvalue of $A$.

LEMMA 2.2. *For any $A \in \mathbb{R}^{m \times r}$, if all $r$ columns of $A$ are linearly independent, then the distance of $x \in \mathbb{R}^m$ to span of $A$ is given by $\|A^\perp x\|^2 = \dfrac{\begin{vmatrix} A^T A & A^T x \\ x^T A & x^T x \end{vmatrix}}{|A^T A|}$.*

*Proof.* Note that by elementary row operations,

$$\begin{vmatrix} A^T A & A^T x \\ x^T A & x^T x \end{vmatrix} = \begin{vmatrix} A^T A & \vdots \\ 0 & x^T x - x^T A(A^T A)^{-1} A^T x \end{vmatrix}$$
$$= |A^T A| \, |x^T A^\perp x| = |A^T A| \, \|A^\perp x\|^2$$

where we used the fact that $A(A^T A)^{-1} A^T = A^\Pi$ and $I - A^\Pi = A^\perp$.

## 3 Bound on ratio of symmetric functions

The following theorem was first proved in the classic paper of Schur [13]. See also [11, Section 3]. We present a different proof below.

THEOREM 3.1. *For any $\sigma \in \mathbb{R}^n \geq 0$, the ratio $\dfrac{\mathbb{S}_{r+1}(\sigma)}{\mathbb{S}_r(\sigma)}$ is Schur-concave.*

*Proof.* By Schur's criterion to establish Schur-concavity of symmetric functions, it suffices to show that

$$\underbrace{\left( \frac{\partial \frac{\mathbb{S}_{r+1}(\sigma)}{\mathbb{S}_r(\sigma)}}{\partial \sigma_i} - \frac{\partial \frac{\mathbb{S}_{r+1}(\sigma)}{\mathbb{S}_r(\sigma)}}{\partial \sigma_j} \right) (\sigma_i - \sigma_j) \leq 0}_{(*)}$$

for all $i, j$. Using the identities

$$\frac{\partial \frac{\mathbb{S}_{r+1}(\sigma)}{\mathbb{S}_r(\sigma)}}{\partial \sigma_i} = \frac{\mathbb{S}_r(\sigma)\mathbb{S}_r(\sigma \setminus \sigma_i) - \mathbb{S}_{r+1}(\sigma)\mathbb{S}_{r-1}(\sigma \setminus \sigma_i)}{\mathbb{S}_r^2(\sigma)}$$
$$\mathbb{S}_k(\sigma \setminus \sigma_i) = \sigma_j \mathbb{S}_{k-1}(\sigma \setminus \{\sigma_i, \sigma_j\}) + \mathbb{S}_k(\sigma \setminus \{\sigma_i, \sigma_j\})$$

we have that

$$(*)\mathbb{S}_r^2(\sigma) = \mathbb{S}_r(\sigma)\left[\mathbb{S}_r(\sigma \setminus \sigma_i) - \mathbb{S}_r(\sigma \setminus \sigma_j)\right]$$
$$\quad - \mathbb{S}_{r+1}(\sigma)\left[\mathbb{S}_{r-1}(\sigma \setminus \sigma_i) - \mathbb{S}_{r-1}(\sigma \setminus \sigma_j)\right]$$
$$= \mathbb{S}_r(\sigma)(\sigma_j - \sigma_i)\mathbb{S}_{r-1}(\sigma \setminus \{\sigma_i, \sigma_j\})$$
$$\quad - \mathbb{S}_{r+1}(\sigma)(\sigma_j - \sigma_i)\mathbb{S}_{r-2}(\sigma \setminus \{\sigma_i, \sigma_j\})$$
$$= (\sigma_j - \sigma_i) \Big[ \mathbb{S}_r(\sigma)\mathbb{S}_{r-1}(\sigma \setminus \{\sigma_i, \sigma_j\})$$
$$\quad - \mathbb{S}_{r+1}(\sigma)\mathbb{S}_{r-2}(\sigma \setminus \{\sigma_i, \sigma_j\}) \Big].$$

Note that if we can show that the expression

$$\mathbb{S}_r(\sigma)\mathbb{S}_{r-1}(\sigma \setminus \{\sigma_i, \sigma_j\}) - \mathbb{S}_{r+1}(\sigma)\mathbb{S}_{r-2}(\sigma \setminus \{\sigma_i, \sigma_j\})$$

is non-negative, we are done. For $r = 2$, $\mathbb{S}_{r-2} = 0$ hence we will consider the case when $r \geq 3$.

We will do so by exhibiting a flow $f$ on a bipartite graph with left nodes labeled with $L = \binom{[n]}{r+1} \times \binom{[n] \setminus \{i,j\}}{r-2}$ and right nodes labeled with $R = \binom{[n]}{r} \times \binom{[n] \setminus \{i,j\}}{r-1}$ with the property that if there is a non-zero flow from $(S, T) \in L$ to $(S', T') \in R$ then $\prod_{i \in S} \sigma_i \prod_{j \in T} \sigma_j \leq \prod_{i \in S'} \sigma_i \prod_{j \in T'} \sigma_j$ and total flow leaving any node on left is 1 whereas total flow entering any node on right is at most 1.

Given $(S, T) \in \binom{[n]}{r+1} \times \binom{[n] \setminus \{i,j\}}{r-2}$, consider $U = S \setminus (T \cup \{i, j\}) \neq \emptyset$. For each $k \in U$, we set

$$f_{(S,T),(S \setminus \{k\}, T \cup \{k\})} = \frac{1}{|U|}.$$

By construction, this satisfies the following:

1. $\sum_{(S',T') \in R} f_{(S,T),(S',T')} = 1$.

2. $f_{(S,T),(S',T')} \left( \prod_{i \in S} \sigma_i \prod_{j \in T} \sigma_j - \prod_{i \in S'} \sigma_i \prod_{j \in T'} \sigma_j \right) = 0$.

In order to prove that $\sum_{(S,T) \in L} f_{(S,T),(S',T')} \leq 1$, if $f_{(S,T),(S',T')} \neq 0$, then there exists $k$ for some $k \in T' \setminus S'$ such that $T = T' \setminus \{k\}$, $S = S' \cup \{k\}$. Hence $|S' \setminus (T' \cup \{i,j\})| = |S \setminus (T \cup \{i,j\})| - 1$. Therefore

$$\sum_{(S,T) \in L} f_{(S,T),(S',T')} = \sum_{k \in T' \setminus S'} \frac{1}{|S' \setminus (T' \cup \{i,j\})| + 1}$$

(3.4)
$$= \frac{|T' \setminus S'|}{|S' \setminus (T' \cup \{i,j\})| + 1}$$

We have $|S'| = |T'| + 1 \geq 3$, $|S' \setminus (T' \cup \{i,j\})| + 1 \geq |S' \setminus T'| - 2 + 1$. Therefore Equation (3.4) can be upper bounded by:

(3.5)
$$\leq \frac{|T' \setminus S'|}{|S' \setminus T'| - 1} = 1$$

where Equation (3.5) follows from $|S'| = |T'| + 1 \implies |S' \setminus T'| = |T' \setminus S'| + 1$.

We now use the Schur-concavity to prove our upper bound on $\frac{\mathbb{S}_{r+1}(\sigma)}{\mathbb{S}_r(\sigma)}$.

LEMMA 3.1. *For any non-negative vector $\rho \in \mathbb{R}^n \geqslant 0$, positive integers $k, r$ such that $r \geqslant k$:*

$$\frac{\mathbb{S}_{r+1}(\rho)}{\mathbb{S}_r(\rho)} \leqslant \frac{1}{r+1-k} \left( \sum_{i \geqslant k+1} \rho\!\downarrow_i \right)$$

*Proof.* Note that, for any $\beta$:

$$\frac{\mathbb{S}_{r+1}(\beta\rho)}{\mathbb{S}_r(\beta\rho)} = \frac{\beta^{r+1}}{\beta^r} \frac{\mathbb{S}_{r+1}(\rho)}{\mathbb{S}_r(\rho)} = \beta \frac{\mathbb{S}_{r+1}(\rho)}{\mathbb{S}_r(\rho)}.$$

Thus without loss of generality, we may assume that $\sum_i \rho_i = 1$. Further, we can assume that $\rho$ is sorted in non-increasing order. Let $\alpha \triangleq \sum_{i \leqslant k} \rho_i$. Consider the following series $\rho'$.

$$\rho'_i = \begin{cases} \frac{1-\alpha}{n-k} & \text{if } i \geqslant k+1, \\ \frac{\alpha}{k} & \text{else.} \end{cases}$$

Since $\rho$ is sorted in non-increasing order, it is easy to see that, for all $i$ we have $\rho'_i \geqslant \rho'_{i+1}$. We have $(\rho'_1, \ldots, \rho'_k) = (\frac{\alpha}{k}, \ldots, \frac{\alpha}{k}) \prec (\rho_1, \ldots, \rho_k)$ and $(\rho'_{k+1}, \ldots, \rho'_n) = (\frac{1-\alpha}{n-k}, \ldots, \frac{1-\alpha}{n-k}) \prec (\rho_{k+1}, \ldots, \rho_n)$. Therefore $\rho' \prec \rho$ which implies:

$$\frac{\mathbb{S}_{r+1}(\rho)}{\mathbb{S}_r(\rho)} \leqslant \frac{\mathbb{S}_{r+1}(\rho')}{\mathbb{S}_r(\rho')}$$

$$= \frac{\sum_{0 \leqslant \ell \leqslant k} \binom{k}{\ell} \binom{n-k}{r-\ell+1} \left(\frac{1-\alpha}{n-k}\right)^{r-\ell+1} \left(\frac{\alpha}{k}\right)^\ell}{\sum_{0 \leqslant \ell \leqslant k} \binom{k}{\ell} \binom{n-k}{r-\ell} \left(\frac{1-\alpha}{n-k}\right)^{r-\ell} \left(\frac{\alpha}{k}\right)^\ell}$$

$$= \frac{1-\alpha}{n-k} \cdot \frac{\sum_{0 \leqslant \ell \leqslant k} \binom{k}{\ell} \frac{n-k-r+\ell}{r-\ell+1} \binom{n-k}{r-\ell} \left(\frac{1-\alpha}{n-k}\right)^{r-\ell} \left(\frac{\alpha}{k}\right)^\ell}{\sum_{0 \leqslant \ell \leqslant k} \binom{k}{\ell} \binom{n-k}{r-\ell} \left(\frac{1-\alpha}{n-k}\right)^{r-\ell} \left(\frac{\alpha}{k}\right)^\ell}$$

$$\leqslant \frac{n-r}{n-k} \frac{1-\alpha}{r-k+1} \leqslant \frac{1}{r-k+1}(1-\alpha) \, .$$

## 4 Bounds on column reconstruction

We now present the upper bound relating the best $r$-column reconstruction of a matrix $X$ to the error $\|X - X_{(k)}\|_F^2$ of the best rank-$k$ approximation in the Frobenius norm.

THEOREM 4.1. *For any $X \in \mathbb{R}^{m \times n}$ and positive integers $r \geqslant k \geqslant 1$,*

$$\min_{S \in \binom{[n]}{r}} \mathsf{Tr}(X^T X_S^\perp X) \leqslant \mathbb{E}_{\mathbf{C} \sim \mathcal{C}_r(X)} \left[ \mathsf{Tr}(X^T X_\mathbf{C}^\perp X) \right]$$

$$\leqslant \frac{r+1}{r+1-k} \|X - X_{(k)}\|^2$$

*where $\mathbf{C} \sim \mathcal{C}_r(X)$ denotes sampling $C$ with probability proportional to determinant of $X_C^T X_C$, $|X_C^T X_C|$. In other words, for any positive real $\varepsilon > 0$,*

$$\min_{S \in \binom{[n]}{k/\varepsilon + k - 1}} \mathsf{Tr}(X^T X_S^\perp X) \leqslant (1+\varepsilon) \|X - X_{(k)}\|^2.$$

*Furthermore, for any $r = o(n)$, this bound is tight up to lower order terms in the number of columns chosen: There exists a matrix $\widetilde{X} \in \mathbb{R}^{n \times n}$ such that*

$$(1 + \varepsilon - o(1)) \|\widetilde{X} - \widetilde{X}_{(k)}\|^2 \leqslant \min_{S \in \binom{[n]}{k/\varepsilon}} \mathsf{Tr}(\widetilde{X}^T \widetilde{X}_S^\perp \widetilde{X}).$$

*Proof.* The first bound is obvious since the minimum is upper bounded by the average. For the second bound, note that $\mathbb{E}_{\mathbf{C} \sim \mathcal{C}_r(X)} \left[ \mathsf{Tr}(X^T X_\mathbf{C}^\perp X) \right]$ is equal to

$$\sum_{S \in \binom{[n]}{r}} \frac{|X_S^T X_S| \mathsf{Tr}(X^T X_S^\perp X)}{\sum_{S \in \binom{[n]}{r}} |X_S^T X_S|}$$

$$= \frac{\sum_S \sum_u |X_S^T X_S| \|X_S^\perp X_u\|^2}{\sum_S |X_S^T X_S|}$$

$$= \frac{\sum_S \sum_u |X_{S,u}^T X_{S,u}|}{\sum_S |X_S^T X_S|} \quad \text{(using Lemma 2.2)}$$

$$= \frac{(r+1) \sum_T |X_T^T X_T|}{\sum_S |X_S^T X_S|}$$

$$= (r+1) \frac{\mathbb{S}_{r+1}(\sigma)}{\mathbb{S}_r(\sigma)} \quad \text{(using Lemma 2.1)}$$

where $\sigma_1 \geqslant \sigma_2 \geqslant \cdots \geqslant \sigma_n \geqslant 0$ are the eigenvalues of $X^T X$. The claimed upper bound now follows by applying the bound from Lemma 3.1 and recalling $\|X - X_{(k)}\|_F^2 = \sum_{i \geqslant k+1} \sigma_i$.

Existence of $\widetilde{X}$ follows from Lemma 6.2 given in Section 6.

## 5 Fast volume sampling algorithm

In this section, we describe and analyze our volume sampling algorithm, which leads to the proof of Theorem 1.2.

THEOREM 5.1. *Given a matrix $X \in \mathbb{R}^{m \times n}$, $m \leqslant n$, and an integer $r$, Algorithm* `Volume-Sample`$(X, r)$ *returns $C \in \binom{[n]}{r}$ with probability $\frac{|X_C^T X_C|}{\sum_{T \in \binom{[n]}{r}} |X_T^T X_T|}$.*

*Furthermore it can be implemented using at most $O(rm^2 n)$ arithmetic operations.*

*Proof of Correctness.* For correctness, notice that for $C$ sampled with probability $|X_C^T X_C|$, if we let $C = \{i_1 <$

**Algorithm 1** Volume-Sample$(X, r)$.

**Input:** $X \in \mathbb{R}^{m \times n}$ and positive integer $r$.
**Output:** $r$ columns of $X$, $C \in \binom{[n]}{r}$, chosen with probability proportional to $|X_C^T X_C|$: $\mathbf{C} \sim \mathcal{C}_r(X)$.
**Procedure:**

1. Let $C \leftarrow \emptyset$. Initialize the table $\mathcal{T}$ of the $n$ outer products $X_{[\ell,n]} X_{[\ell,n]}^T$, $\ell \in [n]$.

2. Choose $\tau$ uniformly at random from $[0, 1]$.

3. $t \leftarrow \tau \cdot \mathbb{S}_r(X^T X)$.

4. For $i \leftarrow 1$ to $r$:

    (a) $\ell \leftarrow 1$, $u \leftarrow n$.
    (b) While $\ell \neq u$
        i. $m \leftarrow \lfloor \frac{\ell+u}{2} \rfloor$.
        ii. $h \leftarrow \mathbb{S}_r\left(X_{[\ell,n]}{}^T X_{[\ell,n]}\right) - \mathbb{S}_r\left(X_{[m+1,n]}{}^T X_{[m+1,n]}\right)$ which is equal to $\mathbb{S}_r\left(X_{[\ell,n]} X_{[\ell,n]}{}^T\right) - \mathbb{S}_r\left(X_{[m+1,n]} X_{[m+1,n]}{}^T\right)$ using $\mathcal{T}$.
        iii. If $t > h$, then $t \leftarrow t - h$, $\ell \leftarrow m + 1$.
        iv. Else $u \leftarrow m$.
    (c) $C \leftarrow C \cup \{\ell\}$, $X \leftarrow X_\ell^\perp X$ and update the table $\mathcal{T}$ of outer products.

5. Return $C$.

---

$i_2 < \ldots < i_r\}$:

$$\Pr_{i_1,\ldots,i_r}\left[i_1 = j\right] = \|X_j\|^2 \frac{\mathbb{S}_{r-1}(X_{[j+1,n]}^T X_j^\perp X_{[j+1,n]})}{\mathbb{S}_r(X^T X)}.$$

Notice that the algorithm, when it exists out of the while loop for the first time, chooses each $\ell$ with probability

$$\frac{\mathbb{S}_r(X_{[\ell,n]}^T X_{[\ell,n]}) - \mathbb{S}_r(X_{[\ell+1,n]}^T X_{[\ell+1,n]})}{\mathbb{S}_r(X^T X)}$$
$$= \|X_\ell\|^2 \frac{\mathbb{S}_{r-1}(X_{[\ell+1,n]}^T X_\ell^\perp X_{[\ell+1,n]})}{\mathbb{S}_r(X^T X)}$$

which completes the proof.

*Proof of Running Time.* We assume each elementary arithmetic operation takes unit time.

By [2, Section 16.6], we can compute $\mathbb{S}_r(X_{[\ell,n]}^T X_{[\ell,n]}) = \mathbb{S}_r(X_{[\ell,n]} X_{[\ell,n]}^T)$ in time $O(m^\omega \log m)$ *given* the outer product $X_{[\ell,n]} X_{[\ell,n]}^T$. Since $X_{A \cup B} X_{A \cup B}{}^T = X_A X_A^T + X_B X_B{}^T$, we can compute the table $\mathcal{T}$ all the $n$ outer products $X_{[\ell,n]} X_{[\ell,n]}^T$, for $\ell \in [n]$, in time $O(m^2 n)$. Also, given $X_\ell$, if we let $z = \frac{X_\ell}{\|X_\ell\|}$:

$$(X_\ell^\perp X_S)(X_\ell^\perp X_S)^T = X_S X_S{}^T + zz^T(z^T X_S X_S^T z)$$
$$- zz^T X_S X_S^T - X_S X_S^T zz^T.$$

Hence, after choosing some column $\ell$, we can update each outer product matrix in the table $\mathcal{T}$ in $O(m^2)$ time. Since there are at most $n$ matrices in this table, each update step takes $O(m^2 n)$ time.

For each column we choose, we evaluate at most $O(\log n)$ many symmetric functions $\mathbb{S}_r$. Thus choosing one column takes time $O(m^\omega \log m \log n)$ given the table $\mathcal{T}$. Since we choose $r$ columns, the total amount of time, including the time to initialize and update $\mathcal{T}$ in each iteration, is bounded by

$$O\left(rm^\omega \log m \log n + rm^2 n\right)$$
$$= O\left(rm^2(m^{\omega-2} \log m \log n + n)\right).$$

Since $m^{\omega-2} \log m \log n \leqslant \sqrt{n} \log^2 n = o(n)$, this bound becomes $O\left(rm^2 n\right)$.

The claim in Theorem 1.2 about the performance of Algorithm Volume-Sample as a column-selection algorithm follows from the upper bound on $\mathbb{E}_{\mathbf{C} \sim \mathcal{C}_r(X)}\left[\mathsf{Tr}(X^T X_{\mathbf{C}}^\perp X)\right]$ in Theorem 4.1.

## 6 Lower bound for column-selection

In this section, we construct matrices for given $k$ and $r$ for which the upper bound stated in Theorem 4.1 is nearly tight. Our construction is in fact the same as the one given by Deshpande and Vempala [6]. Our analysis is different and shows a lower bound on the quantity $\mathsf{Tr}(X^T X_S^\perp X)$ where the full column span of the chosen $r$ columns is allowed for approximating $X$.

DEFINITION 6.1. *Given $\delta > 0$ and $m$, we define $M^{(m,\delta)} \in \mathbb{R}^{m \times m}$ as*

$$M^{(m,\delta)} \triangleq \delta I + J,$$

*where $I$ is the identity matrix of dimension $m$, and $J$ the all 1's $m \times m$ matrix.*

OBSERVATION 6.1. *Given any $\delta > 0$ and positive integer $m$, the followings hold for the matrix $M^{(m,\delta)}$:*

1. $\mathsf{Tr}(M^{(m,\delta)}) = m(1+\delta)$.

2. *Its largest eigenvector is the all 1's vector, with corresponding eigenvalue $\sigma_1 = \sigma_1\left(M^{(m,\delta)}\right)$ given by $\sigma_1 = \delta + m$. Rest of the eigenvalues are all equal with value $\sigma_2 = \sigma_3 = \ldots = \sigma_m = \delta$.*

3. $\left|M^{(m,\delta)}\right| = \prod_{i=1}^m \sigma_i = \delta^m + m\delta^{m-1}$.

LEMMA 6.1. *Given any $\delta > 0$ and positive integer $r$, for $n \geqslant r$, if we let $X^T X = M^{(n,\delta)}$, then*

$$\min_{S \in \binom{[n]}{r}} \frac{\mathsf{Tr}\left(X^T X_S^\perp X\right)}{\|X - X_{(1)}\|_F^2} \geqslant 1 + \frac{k}{r} - o(1).$$

*Proof.* Note that $\|X - X_{(1)}\|_F^2 = \sum_{i \geqslant 2} \sigma_i = (n-1)\delta$. For any subset $C \subseteq [n]$ of size $|C| = r$, the corresponding minor of $X^T X$ is given by

$$X_C^T X_C = M^{(|C|,\delta)} \implies |X_C^T X_C| = \delta^r + r\delta^{r-1}.$$

Consequently for $i \notin C$,

$$\|X_C^\perp X_i\|^2 = \frac{\left|X_{C \cup \{i\}}^T X_{C \cup \{i\}}\right|}{|X_C^T X_C|}$$
$$= \frac{\delta^r(\delta + (r+1))}{\delta^{r-1}(\delta + r)} = \delta\left(1 + \frac{1}{r+\delta}\right).$$

In particular,

$$\mathsf{Tr}(X^T X_C^\perp X) = (n-r)\delta\left(1 + \frac{1}{r+\delta}\right).$$

Therefore

$$\frac{\mathsf{Tr}\left(X^T X_S^\perp X\right)}{\|X - X_{(1)}\|_F^2} = \frac{n-r}{n-1}\left(1 + \frac{1}{r+\delta}\right).$$

LEMMA 6.2. *For any positive integer $n$ and positive integers $k$ and $r$, $r \geqslant k$, such that $r = o(n)$, there exists a matrix of size $n \times n$, $X \in \mathbb{R}^{n \times n}$ for which the following holds:*

$$\min_{S \in \binom{[n]}{r}} \frac{\mathsf{Tr}(X^T X_S^\perp X)}{\|X - X_{(k)}\|^2} \geqslant \frac{n-r}{n-k}\left(1 + \frac{k}{r} - o(1)\right).$$

*Proof.* We will fix $\delta$ to be an infinitesimally small number, $\delta = o(1)$.

For $n = n_0 \cdot k$ with $n_0 \geqslant r+1$, let $X$ be chosen so that $X^T X$ is *block diagonal* matrix of size $n \times n = n_0 k \times n_0 k$ with $k$ copies of $M^{(n_0,\delta)}$ on its diagonals:

$$X^T X = \begin{pmatrix} M^{(n_0,\delta)} & 0^{(n_0)} & \cdots & 0^{(n_0)} \\ 0^{(n_0)} & M^{(n_0,\delta)} & & \vdots \\ \vdots & & \ddots & \\ 0^{(n_0)} & \cdots & & M^{(n_0,\delta)} \end{pmatrix}$$
$$= I^{(k)} \otimes M^{(n_0,\delta)}$$

where we used $0^{(m)}$ and $I^{(m)}$ to denote matrices of size $m \times m$ consisting of all zeroes and identity respectively. Here $\otimes$ denotes tensor (Kronecker) product. By property of tensoring [9], $X^T X$ has $k$ copies of each eigenvalue of $M^{(n_0,\delta)}$. In particular,

(6.6) $\quad \|X - X_{(k)}\|^2 = n(1+\delta) - n - k\delta = (n-k)\delta.$

We will use $[k] \times [n_0]$ to index the columns of matrix $X$, so that for any $i \in [k]$, if we let $X^{(i)} \triangleq X_{\{i\} \times [n_0]}$, we have $X^{(i)T} X^{(i)} = M^{(n_0,\delta)}$, and for any $i \neq j \in [k]$, $X^{(i)T} X^{(j)} = 0^{(n_0)}$.

Proceeding as in [6], given $S$, let $S_i$ be the set of columns chosen from $i^{th}$ block, so that $S_i \triangleq \{j \in [n_0] \mid (i,j) \in S\}$. It is easy to see that,

$$\mathsf{Tr}\left(X^{(i)T} X_S^\perp X^{(i)}\right) = \mathsf{Tr}\left(X^{(i)T} X_{S_i}^{(i)\perp} X^{(i)}\right)$$
$$\geqslant \delta(n_0 - |S_i|)\left(1 + \frac{1}{\delta + |S_i|}\right),$$

where we used Lemma 6.1. Therefore

$$\mathsf{Tr}\left(X^T X_S^\perp X\right) = \sum_i \mathsf{Tr}\left(X^{(i)T} X_{S_i}^{(i)\perp} X^{(i)}\right)$$
(6.7) $$= \sum_i \delta(n_0 - |S_i|)\left(1 + \frac{1}{\delta + |S_i|}\right).$$

Note that $(n-x)(1 + 1/(\delta+x))$ is convex as long as $x + \delta \geqslant 0$. Therefore we can use Jensen's inequality and

lower bound the expression in (6.7) by

$$\delta k \left(n_0 - \frac{1}{k}\sum_i |S_i|\right)\left(1 + \frac{1}{\delta + \frac{1}{k}\sum_i |S_i|}\right)$$
$$=\delta k \left(n_0 - \frac{r}{k}\right)\left(1 + \frac{1}{\delta + \frac{r}{k}}\right)$$
$$=\delta (n-r)\left(1 + \frac{1}{\delta + \frac{r}{k}}\right).$$

Recalling the bound (6.6) for the best rank-$k$ approximation, we see that for any $S$ with $|S| = r = o(n)$ and $\delta = o(1)$:

$$\frac{\mathsf{Tr}\left(X^T X_S^{\perp} X\right)}{\|X - X_{(k)}\|^2} \geqslant \frac{n-r}{n-k}\left(1 + \frac{k}{r}(1-o(1))\right)$$
$$\geqslant 1 + \frac{k}{r} - o(1).$$

## Acknowledgments

We thank Malik Magdon-Ismail for pointing out an inaccuracy in our interpretation of the lower bound in [6].